# High-quality quantum process tomography of time-bin qubit's transmission over a metropolitan fiber network and its application

**Peiyu Zhang(张佩瑜)[1,※], Liangliang Lu(陆亮亮)[1,※], Fangchao Qu(渠方超)[1], Xinhe Jiang(蒋新贺)[1], Xiaodong Zheng(郑晓冬)[1], Yanqing Lu(陆延青)[1], Shining Zhu(祝世宁)[1], Xiao-Song Ma(马小松)[1*]**

[1]National Laboratory of Solid-state Microstructures, School of Physics, College of Engineering and Applied Sciences,
Collaborative Innovation Center of Advanced Microstructures, Nanjing University, 210093 Nanjing, China.

※ These authors contributed equally to this work

*Corresponding author: xiaosong.ma@nju.edu.cn

We employ quantum state and process tomography with time-bin qubits to benchmark a city-wide metropolitan quantum communication system. Over this network, we implement real-time feedback control systems for stabilizing the phase of the time-bin qubits, and obtain a 99.3% quantum process fidelity to the ideal channel, indicating the high quality of the whole quantum communication system. This allows us to implement field trial of high performance quantum key distribution using coherent one way protocol with average quantum bit error rate and visibility of 0.25% and 99.2% during 12 hours over 61 km. Our results pave the way for the high-performance quantum network with metropolitan fibers.

Quantum internet connects quantum computers with quantum communication channels [1, 2], facilitating the transmission of information carried by qubits. Recently, free-space quantum communication has had tremendous advancement [3]. On the other hand, fiber-based quantum communication is a natural candidate for the realization of transmitting quantum information in metropolitan scale. This is because its compatibility with established fiber network for classical communication [4-12].

To obtain the full knowledge of the transmission process over the fiber channel is quintessential for the security and reliability of quantum communication systems. A method for reconstructing the quantum process is known as quantum process tomography (QPT) [13]. Based on the methods, we can fully describe the channels and understand the possible errors during transmission [14-16]. A time-bin qubit is a promising quantum information carrier over fiber networks (e.g., intercity quantum teleportation [17, 18], and quantum key distribution (QKD) [19-21]) because it is easy to prepare, polarization independent and stable in the fiber. However, to the best of our knowledge, there are no tests of QPT in a fiber network based on time-bin qubits encoded in weak coherent states, let alone in an installed metropolitan telecommunication fiber network[14, 15, 22-24]. Here we carry out tomographic protocols based on time-bin encoding to characterize an installed commercial fiber network between the two campuses of Nanjing University. The physical distance between the two campuses is about 30.5 km. We use a fiber loop (about 61 km in total with a loss of 28.02 dB) to guide the photon back to Gulou campus at Nanjing University. By doing so, we double the attenuations of the signal, which enables us to characterize our QKD system under various operating conditions and provides important metrics of our system with high-transmission loss. Full reconstruction of the channel helps us better understand the channel conditions. To verify the reliability of the QPT experiment, we then implement a field trial of coherent one way (COW) QKD [20] with continuous and autonomous feedback control over 12 hours. We obtain the averaged quantum bit error rates (QBERs) of 0.25% and visibilities of 99.2% respectively, matching well with the QPT results. Such technique can be used as a standardized method for the calibration of quantum fiber networks in the future. The COW protocol can be naturally extended to three-state protocol for considering the coherent attacks, which has been studied both theoretically and experimentally [25-27].

An aerial map of the Nanjing University quantum network, identifying the locations of Alice and Bob, is shown in Fig.1 with the schematics of the experimental setup as the insets. There are three nodes in the network with two nodes (node A&B) at Gulou Campus and one node (node C) at Xianlin Campus. These nodes are separated by distances between 0.2km and 30.5km respectively, where the superconducting nanowire single-photon detector (SSPD) is situated at node B. The sender (Alice, at node C for remote experiment and at node A for looped back field trials) prepares time-bin encoded weak laser pulses and sends them to the receiver (Bob) at node A over the commercial fibers, considered as a quantum channel here. On Alice's side, we use a continuous-wave laser at 1536.61 nm (ITU-T channel 51) with intensity modulator (IM) to generate the time-encoded pulses. The pulse width is about 1.5 ns and the pulse separation is about 5 ns. Then, light is sent through an IM bias controller to lock the operating point of IM and ensure a stable operation over time and environmental conditions. The subsequent phase modulator (PM) applies a relative phase between two pulses. The attenuator is introduced to reduce the average photon number per pulse. The system is synchronized via co-propagating multiplexed pulses in different wavelength (1548.51 nm, ITU-T channel 36) and polarization with respect to the quantum signals at a rate of 1 MHz. At Alice's side, the arbitrary wave generator (AWG) is used to drive the PM for quantum signal and IMs for quantum and synchronization signals, respectively. At Bob's side, a dense wavelength division multiplexer (DWDM) filter is used to separate the classical and quantum signals. The polarization beam splitter (PBS) and electrical polarization controller (EPC) are used to perform polarization stabilization. For QPT, Bob projects the incoming photon onto the standard Pauli bases at setup Bob2, which are realized by two Faraday-Michelson interferometers (FMIs) with a 1m difference in the two arms [21], where a phase shifter (PS) is employed to determine the relative phase information of the two time bins. For COW QKD, Bob can decode the qubits at setup Bob1 with a 90:10 beam splitter to passively route most of the photons for arrival time measurements. The remaining 10% are fed into an FMI for measuring the phase coherence. Photons are then transmitted to node B and detected by superconducting single-photon detectors (SSPDs) with 80% detection efficiency, the corresponding electronic signals return to node A through coaxial cables and are collected by the field programmable gate array (FPGA) with 156 ps resolution. Note that three-state protocol can also be implemented in this setup [26]. To optimize the visibility, we develop a real-time proportional-integral-derivative (PID) feedback system, where a thermal phase shifter (PS) is used to compensate the phase drifts of the interferometer per 0.47 seconds with the error count rate in monitor line as the feedback. The mean photon number of 0.29 per pulse, which is optimized by considering the measured transmission loss and the detection efficiency, according to the security proof by Branciard et al. [28].

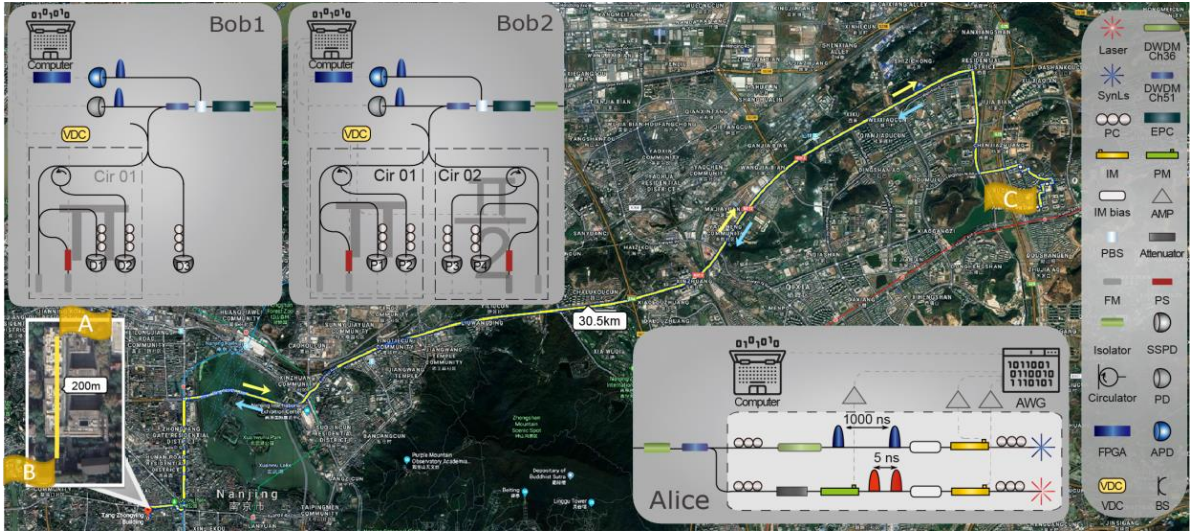

**Fig. 1.** Schematics of the experimental set-up in Nanjing University optical fiber network. Node A and node B are located in Zhongying Tang building and Electron Microscope building respectively in Gulou Campus. Node C is located in fundamental laboratory building in Xianlin Campus. These nodes are separated by distances between 0.2 km and 30.5 km. Fiber installed along the yellow line. Abbreviations of components: IM, intensity modulator; IM bias, intensity modulator bias; AMP, amplifier; PM, phase modulator; PBS, polarization beam splitter; BS, beam splitter; PC, polarization controller; EPC, electrical polarization controller; DWDM, dense wavelength division multiplexer; SynLs, synchronized laser; FM, faraday rotation mirror; PS, phase shifter; SSPD, superconducting single-photon detector; PD, power detector; APD, avalanche photodiode FPGA, field programmable gate array; VDC, variable direct current. Imagery©2020 Google. Map data from Google, Maxar Technologies, CNES/Airbus.

To characterize the performance of the quantum system, we perform single-qubit quantum state tomography (QST) on the quantum states transmitted over the 61.1 km looped back fiber. We create photons in, and project them onto, well-defined time-bin states, such as $|0\rangle$, $|1\rangle$, $|\pm\rangle = 1/\sqrt{2}(|0\rangle \pm |1\rangle)$ and $|\pm i\rangle = 1/\sqrt{2}(|0\rangle \pm i|1\rangle)$, where $|0\rangle$ ($|1\rangle$) stands for the quantum state of photon being early (late) temporal mode. The density matrices of the six final output states reconstructed by QST are shown in Fig. 2(a) in green bars, which are very close to ideal states (black line).

Figure 2(b) shows the state fidelities, which is defined as the overlap between the ideal states and the final output states. The fidelities for the six states are estimated to be $0.997429 \pm 0.000006$ ($|0\rangle$), $0.998614 \pm 0.000004$ ($|1\rangle$), $0.9944 \pm 0.0007$ ($|+\rangle$), $0.9962 \pm 0.0006$ ($|-\rangle$), $0.9957 \pm 0.0006$ ($|+i\rangle$) and $0.9940 \pm 0.0007$ ($|-i\rangle$), respectively. The process of transmitting qubits over this quantum channel is quantified by QPT. We choose $|0\rangle$, $|1\rangle$, $|+\rangle$ and $|+i\rangle$ as the input states $\rho_{in}$ and their corresponding output states $\rho_{out}$ to determine the process matrix $\chi$. The output states are related to the input states through the process density matrix, i.e. $\rho_{out} = \sum_{l,k=0}^{3} \chi_{lk} \sigma_l \rho_{in} \sigma_k$, where $\sigma_{l(k)}$ are the Pauli matrices with $\sigma_0$ being the identity operator. The real and imaginary parts of $\chi$ are shown Fig. 2(c) and (d). The process fidelity is defined as $F_{proc} = Tr(\chi_{ideal}\chi)$, where $\chi_{ideal}$ is the ideal process matrix. There are three fidelities: the fidelity of the output state to the ideal state (F0), the fidelity of the input state to the ideal state (F1) and the fidelity of the output state to the input state (F2). In our work, we obtain the following: F0=99.3%±0.7%, F1=99.42%±0.68% and F2=98.86%±0.63%. All three numbers agree each other within the statistics. The $\sigma_x$, $\sigma_y$ and $\sigma_z$ components of the matrix $\chi$ represent the probabilities of a bit-flip or phase-flip error in the channel. A single-qubit quantum process can be represented graphically subjected to the quantum process [16]. In Fig. 2(e), we plot the ideal states as a wire grid of the Bloch sphere. After the long-fiber transmission from Alice and Bob, the receiving quantum states are, although very close to, not the same as the original states. Therefore, the ideal Bloch sphere is deformed into a slightly anisotropic ellipsoid as shown in the solid blue color.

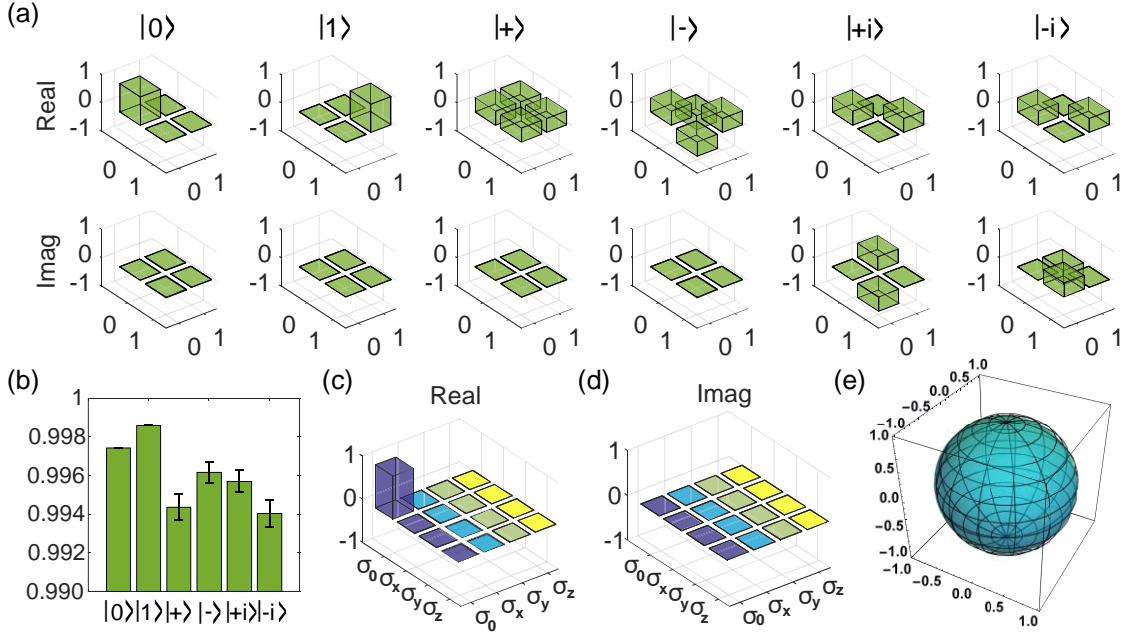

**Fig. 2.** Characterization of the quantum channel. (a) Density matrices of output time-bin encoded states. (b) State fidelities of the six output states to the ideal states. (c)-(d) Real and imaginary parts of the process matrices for the quantum channel, with a fidelity of F0=99.3%±0.7%. (e) Bloch sphere representation of the process. The plot shows how the ideal states on the surface of the Bloch sphere (meshed) are influenced by the quantum channel, with the output states lying on the solid surface. The uncertainties in state fidelities are calculated using a Monte Carlo routine assuming Poissonian errors.

Having established this high-quality quantum system, we proceed to perform QKD by employing COW protocol [20, 28-34]. The coherent pulses chopped by Alice are either empty or have a mean photon number μ=0.29. Each logical bit of information is defined by the position of non-empty pulse in neighboring bins, for example μ-0 for a logical "0" or 0-μ for a logical "1". Decoy sequences μ-μ are sent to prevent photon-number-splitting attacks[28]. To obtain the key, Bob measures the arrival time of the photons on his data line, detector Ds in Bob1 of Fig. 1. In order to avoid Raman noise generated by synchronized signals, along with them we send empty sequences which are not used for coding. Attenuated laser pulses with 1.5 ns width are modulated to signal, decoy and empty sequences respectively. Among them, we send decoy sequences with a probability of 7%, which are sufficient to calibrate the phase drifts during the PID feedback time as well as detect the presence of an eavesdropper. In order to avoid Raman noise generated by synchronized signals, along with them we send empty sequences with a probability of 3%, which is not used for coding. The remaining 90% of the sequences are encoded as signal to obtain a high key rate. To ensure the security, Bob randomly measures the coherence between successive non-empty pulses, such as bit sequences "1-0" or decoy sequences, with the unbalanced interferometer and detectors D1 and D2. Ideally, due to the coherence between pulses, we have all detections on D1 and no detection on D2. A loss of coherence, hence reduced visibility, indicates the presence of disturbance, in which case the key is simply discarded. Coherence can be quantified by the visibility of the interference

$$V = \frac{c(D1)-c(D2)}{c(D1)+c(D2)}, \quad (1)$$

where $c(D1)$ and $c(D2)$ are respectively the detector counts of D1 and D2. It is crucial to investigate the system stability under different operating conditions, such as temperature, time and so on. We summarize this information in Table 1, which includes the length of fibers, the date, the measurement time and the corresponding environment temperature.

Table 1. Characteristics of our system under test

| Fiber links | Length (km) | Total attenuation (dB) | Date (2019) | Time | Temperature |
|---|---|---|---|---|---|
| A-B | 0.2 | 12.95 | 10pm, Jun. 09 - 10am, Jun. 10 | 12h | 22℃-31℃ |
| B-C | 30.5 | | | | |

Fig. 3 shows the quantum bit error rate (QBER) and the interference visibility over a 61.1 km looped back field trials for 12 hours. The averaged QBER and visibility of the system are $0.250\% \pm 0.006\%$ and $0.992 \pm 0.002$, respectively, indicating the high performance of our system. These results match state fidelities well, which proves the reliability and accuracy of our QPT method. The figure also illustrates the system's long-time continuous operation capability. Moreover, from the interference visibility, a phase error rate of about 0.004 can be expected during the key exchange scenario, which is low even compared with other indoor QKD protocols at similar attenuations [34-37].

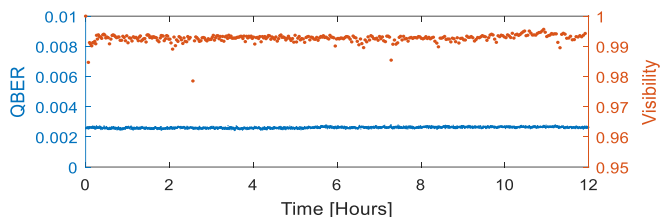

**Fig. 3.** A 12hours continuous operation of Nanjing quantum network with excellent system parameters: quantum bit error rate (blue, left vertical axis) and (red, right vertical axis) for COW protocol.

Fig. 4 shows the secure key rate (SKR) per pulse as a function of channel attenuation. The error correction efficiency is set to 1.16. The SKRs of 30.5 km ($5.78*10^{-4}$ bit/pulse, about 115.6 Kbits/s, for 12.95 dB loss) and 61.1 km ($1.82*10^{-5}$ bit/pulse, about 3.64 Kbits/s, for 28.02 dB loss), marked as red and green pentagrams respectively, are estimated using the above system parameters with the measured QBER and visibilities. According to the security proof by Branciard et al. [28], it has been shown to be an upper bound under the assumption of collective attacks (i.e., Eve interacts with each individual state using the same strategy). We calculate the key rate in the infinite key scenario. As the channel attenuation increases, the number of counts decreases and the dark count rates (DCRs) of the SSPDs (about $10^{-7}$/ns) becomes a major component of QBER, thus the secure key rate decreases exponentially. With the high visibility and negligible DCRs, our system can tolerate more channel loss, which means a wider area network.

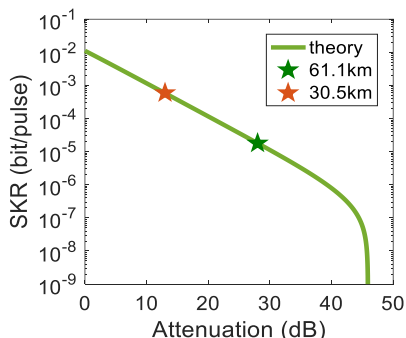

**Fig. 4.** Field trial SKR as a function of attenuation. Green and red pentagrams are our SKR on the network.

With these field tests of our network for quantum communications, we have fully evaluated the quality of the system via both quantum state and process tomography techniques. The QPT technique can be a standardized method for calibrating the quantum fiber networks in future. We have extended a high security key rate per pulse for COW protocol over the installed commercial fiber network with a real-time feedback control. Our results pave the way for the high-performance quantum network with metropolitan fibers.

The authors thank H. Hu, M. Zhou and S. Wu for providing access to the fiber link between the Nanjing University Gulou campus to Nanjing University Xianlin campus, J. Wan for providing laboratory space. This research is supported by the National Key Research and Development Program of China (2017YFA0303704, 2019YFA0308704), National Natural Science Foundation of China (Grant No. 11674170, 11690032, 11321063, 11804153), NSFC-BRICS (No. 61961146001), NSF Jiangsu Province (No. BK20170010), the program for Innovative Talents and Entrepreneur in Jiangsu, and the Fundamental Research Funds for the Central Universities.

## References
1. H. J. Kimble, "The quantum internet," Nature **453,** 1023 (2008).
2. S. Wehner, D. Elkouss, and R. Hanson, "Quantum internet: A vision for the road ahead," Science **362,** eaam9288 (2018).
3. S. Liao, W. Cai, W. Liu, L. Zhang, Y. Li, J. Ren, J. Yin, Q. Shen, Y. Cao, Z. Li, F. Li, X. Chen, L. Sun, J. Jia, J. Wu, X. Jiang, J. Wang, Y. Huang, Q. Wang, Y. Zhou, L. Deng, T. Xi, L. Ma, T. Hu, Q. Zhang, Y. Chen, N. Liu, X. Wang, Z. Zhu, C. Lu, R. Shu, C. Peng, J. Wang, and J. Pan, "Satellite-to-ground quantum key distribution," Nature **549,** 43 (2017).
4. J. Qiu, "Quantum communications leap out of the lab," Nature **508,** 441 (2014).
5. D. Stucki, M. Legre, F. Buntschu, B. Clausen, N. Felber, N. Gisin, L. Henzen, P. Junod, G. Litzistorf, P. Monbaron, L. Monat, J. Page, D. Perroud, G. Ribordy, A. Rochas, S. Robyr, J. Tavares, R. Thew, P. Trinkler, S. Ventura, R. Voirol, N. Walenta, and H. Zbinden, "Long-term performance of the SwissQuantum quantum key distribution network in a field environment," New J. Phys. **13,** 123001 (2011).
6. M. Sasaki, M. Fujiwara, H. Ishizuka, W. Klaus, K. Wakui, M. Takeoka, S. Miki, T. Yamashita, Z. Wang, A. Tanaka, K. Yoshino, Y. Nambu, S. Takahashi, A. Tajima, A. Tomita, T. Domeki, T. Hasegawa, Y. Sakai, H. Kobayashi, T. Asai, K. Shimizu, T. Tokura, T. Tsurumaru, M. Matsui, T. Honjo, K. Tamaki, H. Takesue, Y. Tokura, J. F. Dynes, A. R. Dixon, A. W. Sharpe, Z. L. Yuan, A. J. Shields, S. Uchikoga, M. Legré, S. Robyr, P. Trinkler, L. Monat, J. Page, G. Ribordy, A. Poppe, A. Allacher, O. Maurhart, T. Länger, M. Peev, and A. Zeilinger, "Field test of quantum key distribution in the Tokyo QKD Network," Opt. Express **19,** 10387 (2011).
7. M. Peev, C. Pacher, R. Alléaume, C. Barreiro, J. Bouda, W. Boxleitner, T. Debuisschert, E. Diamanti, M. Dianati, J. F. Dynes, S. Fasel, S. Fossier, M. Fürst, J. Gautier, O. Gay, N. Gisin, P. Grangier, A. Happe, Y. Hasani, M. Hentschel, H. Hübel, G. Humer, T. Länger, M. Legré, R. Lieger, J. Lodewyck, T. Lorünser, N. Lütkenhaus, A. Marhold, T. Matyus, O. Maurhart, L. Monat, S. Nauerth, J. Page, A. Poppe, E. Querasser, G. Ribordy, S. Robyr, L.

Salvail, A. W. Sharpe, A. J. Shields, D. Stucki, M. Suda, C. Tamas, T. Themel, R. T. Thew, Y. Thoma, A. Treiber, P. Trinkler, R. Tualle-Brouri, F. Vannel, N. Walenta, H. Weier, H. Weinfurter, I. Wimberger, Z. L. Yuan, H. Zbinden, and A. Zeilinger, "The SECOQC quantum key distribution network in Vienna," New J. Phys. **11,** 075001 (2009).
8. C. Elliott, "Building the quantum network," New J. Phys. **4,** 46 (2002).
9. Y. Mao, B. Wang, C. Zhao, G. Wang, R. Wang, H. Wang, F. Zhou, J. Nie, Q. Chen, Y. Zhao, Q. Zhang, J. Zhang, T. Chen, and J. Pan, "Integrating quantum key distribution with classical communications in backbone fiber network," Opt. Express **26,** 6010 (2018).
10. B. Fröhlich, J. F. Dynes, M. Lucamarini, A. W. Sharpe, Z. Yuan, and A. J. Shields, "A quantum access network," Nature **501,** 69 (2013).
11. W. Chen, Z. Han, T. Zhang, H. Wen, Z. Yin, F. Xu, Q. Wu, Y. Liu, Y. Zhang, X. Mo, Y. Gui, G. Wei, and G. Guo, "Field experiment on a "star type" metropolitan quantum key distribution network," IEEE Photon. Technol. Lett. **21,** 575 (2009).
12. X. Wang, S. Li, X. Jiang, J. Hu, M. Xue, S. Xu, and S. Pan, "High-accuracy optical time delay measurement in fiber link," Chin. Opt. Lett. **17,** 060601 (2019).
13. M. A. Nielsen and I. L. Chuang, *Quantum Computation and Quantum Information* (Cambridge University Press, 2010).
14. F. Bouchard, F. Hufnagel, D. Koutný, A. Abbas, A. Sit, K. Heshami, R. Fickler, and E. Karimi, "Quantum process tomography of a high-dimensional quantum communication channel," Quantum **3,** 138 (2019).
15. B. Ndagano, B. Perez-Garcia, F. S. Roux, M. McLaren, C. Rosales-Guzman, Y. Zhang, O. Mouane, R. I. Hernandez-Aranda, T. Konrad, and A. Forbes, "Characterizing quantum channels with non-separable states of classical light," Nat. Phys. **13,** 397 (2017).
16. N. Gisin, G. Ribordy, W. Tittel, and H. Zbinden, "Quantum cryptography," Rev. Mod. Phys **74,** 145 (2002).
17. R. Valivarthi, M. l. G. Puigibert, Q. Zhou, G. H. Aguilar, V. B. Verma, F. Marsili, M. D. Shaw, S. W. Nam, D. Oblak, and W. Tittel, "Quantum teleportation across a metropolitan fibre network," Nat. Photonics **10,** 676 (2016).
18. Q. Sun, Y. Mao, S. Chen, W. Zhang, Y. Jiang, Y. Zhang, W. Zhang, S. Miki, T. Yamashita, H. Terai, X. Jiang, T. Chen, L. You, X. Chen, Z. Wang, J. Fan, Q. Zhang, and J. Pan, "Quantum teleportation with independent sources and prior entanglement distribution over a network," Nat. Photonics **10,** 671 (2016).
19. A. Boaron, G. Boso, D. Rusca, C. Vulliez, C. Autebert, M. Caloz, M. Perrenoud, G. Gras, F. Bussières, M. Li, D. Nolan, A. Martin, and H. Zbinden, "Secure quantum key distribution over 421 km of optical fiber," Phys. Rev. Lett. **121,** 190502 (2018).
20. D. Stucki, N. Brunner, N. Gisin, V. Scarani, and H. Zbinden, "Fast and simple one-way quantum key distribution," Appl. Phys. Lett. **87,** 194108 (2005).
21. X. Song, H. Li, C. Zhang, D. Wang, S. Wang, Z. Yin, W. Chen and Z. Han, "Analysis of Faraday-Michelson quantum key distribution with unbalanced attenuation,"Chin. Opt. Lett. **13**, 012701 (2015).
22. W. Liang, S. Wang, H. Li, Z. Yin, W. Chen, Y. Yao, J. Huang, G. Guo, and Z. Han, "Proof-of-principle experiment of reference-frame-independent quantum key distribution with phase coding," Sci. Rep. **4,** 3617 (2014).
23. H. Takesue and Y. Noguchi, "Implementation of quantum state tomography for time-bin entangled photon pairs," Opt. Express **17,** 10976 (2009).
24. D. Bruß, "Optimal eavesdropping in quantum cryptography with six states," Phys. Rev. Lett. **81,** 3018 (1998).
25. D. Rusca, A. Boaron, F. Grünefelder, A. Martin, and H. Zbinden, "Finite-key analysis on the 1-decoy state QKD protocol," Appl. Phys. Lett. **112,** 171104 (2018).
26. D. Bacco, I. Vagniluca, B. Da Lio, N. Biagi, A. Della Frera, D. Calonico, C. Toninelli, F. S. Cataliotti, M. Bellini, L. K. Oxenløwe, and A. Zavatta, "Field trial of a three-state quantum key distribution scheme in the Florence metropolitan area," EPJ Quantum Technol. **6,** 5 (2019).
27. D. Bacco, B. Da Lio, D. Cozzolino, F. Da Ros, X. Guo, Y. Ding, Y. Sasaki, K. Aikawa, S. Miki, H. Terai, T. Yamashita, J. S. Neergaard-Nielsen, M. Galili, K. Rottwitt, U. L. Andersen, T. Morioka, and L. K. Oxenløwe, "Boosting the secret key rate in a shared quantum and classical fibre communication system," Communications Physics **2,** 140 (2019).
28. C. Branciard, N. Gisin, and V. Scarani, "Upper bounds for the security of two distributed-phase reference protocols of quantum cryptography," New J. Phys. **10,** 013031 (2008).
29. X. Tang, A. Wonfor, R. Kumar, R. V. Penty, and I. H. White, "Quantum-safe metro network with low-latency reconfigurable quantum key distribution," J. Lightwave Technol. **36,** 5230 (2018).
30. P. Sibson, C. Erven, M. Godfrey, S. Miki, T. Yamashita, M. Fujiwara, M. Sasaki, H. Terai, M. G. Tanner, C. M. Natarajan, R. H. Hadfield, J. L. O'Brien, and M. G. Thompson, "Chip-based quantum key distribution," Nat. Commun. **8,** 13984 (2017).
31. B. Korzh, C. C. W. Lim, R. Houlmann, N. Gisin, M. J. Li, D. Nolan, B. Sanguinetti, R. Thew, and H. Zbinden, "Provably secure and practical quantum key distribution over 307 km of optical fibre," Nat. Photonics **9,** 163 (2015).
32. D. Stucki, N. Walenta, F. Vannel, R. T. Thew, N. Gisin, H. Zbinden, S. Gray, C. Towery, and S. Ten, "High rate, long-distance quantum key distribution over 250 km of ultra low loss fibres," New J. Phys. **11,** 075003 (2009).
33. D. Stucki, C. Barreiro, S. Fasel, J. Gautier, O. Gay, N. Gisin, R. Thew, Y. Thoma, P. Trinkler, F. Vannel, and H. Zbinden, "Continuous high speed coherent one-way quantum key distribution," Opt. Express **17,** 13326 (2009).
34. G. L. Roberts, M. Lucamarini, J. F. Dynes, S. J. Savory, Z. Yuan, and A. J. Shields, "Modulator-Free Coherent-One-Way Quantum Key Distribution," Laser Photonics Rev. **11,** 1700067 (2017).
35. A. V. Gleim, V. I. Egorov, Y. V. Nazarov, S. V. Smirnov, V. V. Chistyakov, O. I. Bannik, A. A. Anisimov, S. M. Kynev, A. E. Ivanova, R. J. Collins, S. A. Kozlov, and G. S. Buller, "Secure polarization-independent subcarrier quantum key distribution in optical fiber channel using BB84 protocol with a strong reference," Opt. Express **24,** 2619 (2016).

36. B. Korzh, N. Walenta, R. Houlmann, and H. Zbinden, "A high-speed multi-protocol quantum key distribution transmitter based on a dual-drive modulator," Opt. Express **21,** 19579 (2013).
37. Z. Yin, Z. Han, W. Chen, F. Xu, Q. Wu, and G. Guo, "Experimental decoy state quantum key distribution over 120 km fibre," Chin. Phys. Lett. **25,** 3547 (2008).